\documentclass[twocolumn,amsfont,amssymb,amsmath, showpacs,balancelastpage, nofootinbib]{revtex4-1}
\pdfoutput=1

\usepackage{graphicx}
\usepackage{dcolumn}
\usepackage{bm}
\usepackage{amssymb,amsmath,bm}  
\usepackage{color}
\usepackage[colorlinks,linkcolor=red,citecolor=blue,urlcolor=blue ]{hyperref}
\usepackage{multirow}
\usepackage[utf8]{inputenc}
\usepackage{balance}
\usepackage{enumitem}
\newcommand{\araa}{ARA\&A}
\newcommand{\nv}{\hat{\bf n}}

\newcommand{\jcap}{JCAP}
\newcommand{\mnras}{MNRAS}
\newcommand{\aap}{A\&A}

\newcommand{\apjs}{ApJS}
\newcommand{\apjl}{ApJL}

\newcommand{\procspie}{Proceedings of the SPIE}

\begin{document}
  \title{Measurement of the thermal Sunyaev-Zel'dovich effect around cosmic voids}
  \author{David Alonso$^1$}\email{david.alonso@physics.ox.ac.uk}
  \author{J. Colin Hill$^{2,3}$}
  \author{Ren\'{e}e Hlo\v{z}ek$^4$}
  \author{David N. Spergel$^{3,5}$}
  \affiliation{$^{1}$University of Oxford, Denys Wilkinson Building, Keble Road,
               Oxford OX1 3RH, UK}
  \affiliation{$^{2}$Department of Astronomy, Columbia University, New York,
               NY 10027, USA}
  \affiliation{$^{3}$Center for Computational Astrophysics, Flatiron Institute,
               162 5th Avenue, 10010, New York, NY, USA}
  \affiliation{$^{4}$Dunlap Institute for Astronomy and Astrophysics \& \\
               Department of Astronomy and Astrophysics, University of Toronto,
               50 St. George Street, ON M5S 3H4, Canada}
  \affiliation{$^{5}$Department of Astrophysical Sciences, Princeton University,
               Peyton Hall, Princeton NJ 08544-0010, USA}

  \begin{abstract}
    We stack maps of the thermal Sunyaev-Zel'dovich effect produced by the \textit{Planck} Collaboration around the centers of cosmic voids defined by the distribution of galaxies in the CMASS sample of the Baryon Oscillation Spectroscopic Survey, scaled by the void effective radii. We report a first detection of the associated cross-correlation at the $3.4\sigma$ level: voids are under-pressured relative to the cosmic mean. We compare the measured Compton-$y$ profile around voids with a model based solely on the spatial modulation of halo abundance with environmental density. The amplitude of the detected signal is marginally lower than predicted by an overall amplitude $\alpha_v=0.67\pm0.2$. We discuss the possible interpretations of this measurement in terms of modelling uncertainties, excess pressure in low-mass halos, or non-local heating mechanisms.
  \end{abstract}

  \maketitle

  \section{Introduction}\label{sec:intro}
    While the primordial fluctuations are very well described by a Gaussian distribution, the gravitational growth of structure produces a skewed distribution of densities with high density peaks and large voids. Voids can be thought of as the gravitational converse of clusters: as structure grows through cosmic time voids become emptier and emptier while overdensities accumulate mass. Voids have many virtues as a a target for cosmological study. Unlike clusters, they are only mildly non-linear. They are dominated by dark energy so are sensitive to its nature \cite{2012MNRAS.426..440B,2016PhRvL.117i1302H,2017PhRvD..95h3502A,2017JCAP...07..014H}. They are also sensitive to cosmic parameters and complement other probes of structure \cite{2012ApJ...754..109L}. Nevertheless, most cosmology studies focus on the galaxies and clusters that make up only a small fraction of the volume of the Universe.  Voids are more difficult to study: luminous sources (e.g., galaxies) are readily available to trace the overdensities, but can only be used to delineate the boundaries of voids. As such, different approaches exist to identify voids \cite{2008MNRAS.387..933C}, some optimized for truly three-dimensional voids and others for projected two-dimensional underdensities. Much effort has been spent studying the density profile of voids as a function of radius from either the void center or its boundary \cite{2014PhRvL.112y1302H,2016MNRAS.457.2540C}.

    Several groups have been studying the mass profiles of voids through stacking lensing convergence or cosmic microwave background (CMB) temperature maps, around the positions of voids identified using galaxy catalogs \cite{2008ApJ...683L..99G,2014MNRAS.440.2922M,2015MNRAS.454.3357C,2016ApJ...830L..19N,2017ApJ...836..156C,2017MNRAS.465..746S}. These studies have mostly revealed mass profiles consistent with theoretical expectation.
    
    In this paper, we study the pressure profile of voids.  We use measurements of the the thermal Sunyaev-Zel'dovich (tSZ) effect \cite{1972CoASP...4..173S}. Caused by the inverse-Compton scattering of cooler CMB photons off hot electrons, the tSZ effect is a frequency-dependent source of secondary anisotropy in the CMB temperature field. This frequency dependence allows it to be disentangled from the primary CMB signal, which is frequency-independent in thermodynamic temperature units. By combining sky maps at different microwave and radio frequencies, one can produce maps of the expected tSZ signal, which generally traces cosmic overdensities, as detailed below. The tSZ signal from galaxies, clusters of galaxies, and quasars has been observed through stacking, cross-correlation, and matched-filter analyses \cite{2013JCAP...07..008H,2015ApJS..216...27B,2016A&A...594A..27P,2011ApJ...736...39H,2011A&A...536A..12P, 2013A&A...557A..52P, 2015ApJ...808..151G,2017arXiv170600395J,2017MNRAS.467.2315V, 2017arXiv170603753H,2016MNRAS.458.1478C, 2015ApJ...802..135R,2016A&A...588A..61V,2016ApJ...819..128S, 2017ApJ...834..102S,2017MNRAS.468..577S}. In this work we will focus on stacking the tSZ signal around the locations of voids, which has heretofore escaped attention.

    A detailed knowledge of the general properties and distribution of gas in different environments is essential for a precise understanding of the physics of structure formation. Although recent measurements of the kinematic SZ effect have demonstrated that the baryon abundance at low redshift is consistent with that inferred from the early Universe via Big Bang Nucleosynthesis and the primary CMB~\cite{2016PhRvL.117e1301H,2016PhRvD..94l3526F,2012PhRvL.109d1101H,2016A&A...586A.140P, 2016PhRvD..93h2002S,2017JCAP...03..008D,2016MNRAS.461.3172S} (thus resolving the long-standing ``missing baryons'' puzzle~\cite{2004ApJ...616..643F,2007ARA&A..45..221B}), constraints on the precise distribution and thermodynamic state of the diffuse gas remain weak.  Theoretical models predict that a substantial fraction of the baryons resides in a warm-hot plasma associated with low-density structures such as filaments, known as the warm-hot intergalactic medium~\cite{2006ApJ...650..560C}. Cross-correlations of the tSZ effect with various tracers of the matter distribution, such as gravitational lensing maps~\cite{2014PhRvD..89b3508V,2014JCAP...02..030H,2015ApJ...812..154B,2015JCAP...10..047H,2015JCAP...09..046M,2017ApJ...845...71A,2017MNRAS.471.1565H}, can be used to probe the pressure content of gas within and beyond the virial radius of halos, thus placing constraints on models of feedback (e.g., from active galactic nuclei) in structure formation.
    
    Since cosmic voids probe the lowest-density environments of the matter distribution, and have the largest volume filling factor of all elements of the cosmic web, their cross-correlation with tSZ maps can be used to probe the presence of hot gas in low-density regions as well as its properties, as a first step towards a more comprehensive study of the gas temperature-density relation. As we show below, standard models predict that the gas in voids should be under-pressured relative to the cosmic mean, because there is a deficit of massive objects in voids, as compared to average-density regions in the Universe.  However, some ``non-local heating'' models can change this prediction~\cite{2012ApJ...752...23C}, even to the point of yielding an inverted density-temperature relation~\cite{2008MNRAS.386.1131B}.  Our analysis is a first step toward testing these scenarios.  Moreover, as the tSZ signal in voids receives a relatively larger contribution from low-mass halos than that in high-density regions, our measurement is also useful for constraining the behavior of the tSZ -- mass relation at low masses.  Recent analyses have found inconclusive results regarding the consistency of this relation with the self-similar prediction ($Y \propto M^{5/3}$) at low masses~\cite{2013A&A...557A..52P,2015ApJ...808..151G,2015MNRAS.451.3868L,2017arXiv170603753H}. Void analyses, such as ours, may be useful in shedding light on this issue using forthcoming datasets.
    
    The paper is structured as follows. Section \ref{sec:theory} describes the model used here to predict the tSZ signal around voids. In Section \ref{sec:data} we summarize the datasets (void catalog and CMB maps) used in the analysis, the results of which are detailed in Section \ref{sec:results}. Finally, Section \ref{sec:discussion} summarizes our findings and discusses their interpretation. Throughout this work we assume a flat $\Lambda$CDM cosmology with parameters $\Omega_M=0.3$, $h=0.7$, $\sigma_8=0.8$, and $n_s=0.96$, where $\Omega_M$ is the fractional density of non-relativistic species today, $h$ is the normalized expansion rate, $\sigma_8$ is the standard deviation of the linear matter overdensity in spheres with a radius of $8\,h^{-1}{\rm Mpc}$ at $z=0$, and $n_s$ is the primordial spectral index of scalar perturbations. The choice of $\Omega_M$ was made to coincide with the value assumed in the construction of the void catalog used in this analysis (see Section \ref{ssec:data.voids}). This was necessary in order to transform the comoving lengths used in the catalog into projected angular separations.
    
  \section{The expected void SZ profile}\label{sec:theory}
\begin{figure}
      \centering
      \includegraphics[width=0.49\textwidth]{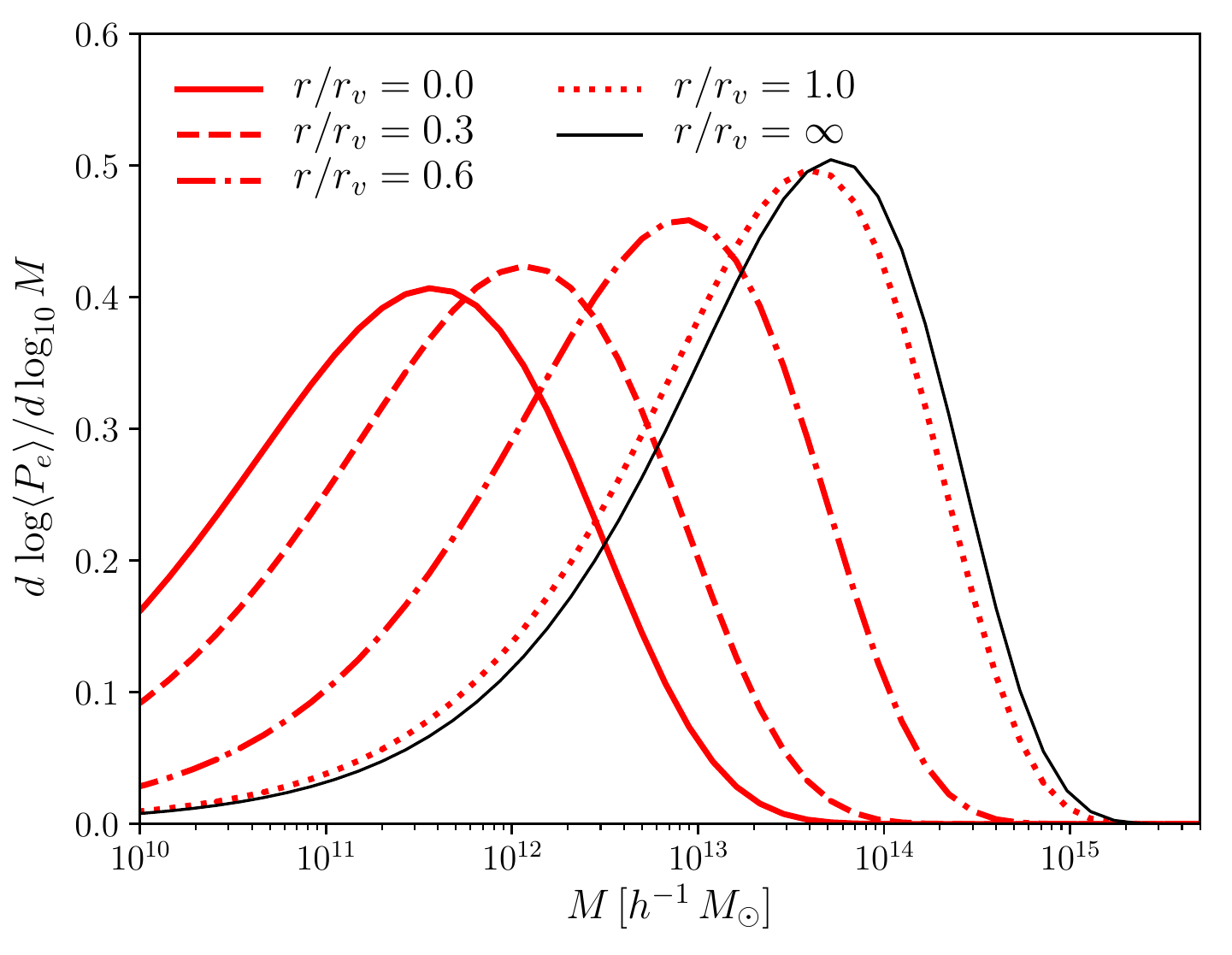}
      \caption{Relative contribution to the average electron pressure from halos of different mass at different distances from the void center (here $r_v$ is the effective void radius). Due to the suppressed growth of structure inside the void, the electron pressure is mostly supported by low-mass halos.}
      \label{fig:pe_mass}
    \end{figure}

    The thermal Sunyaev-Zel'dovich effect \cite{1972CoASP...4..173S} traces the hot gas in the Universe through the inverse Compton scattering of CMB photons by high-energy electrons. This induces a spatial and spectral distortion in the CMB given by
    \begin{equation}
      \frac{\Delta T(\nv)}{T_{\rm CMB}}=g\left(\frac{h\nu}{k_B\,T_{\rm CMB}}\right)\,y(\nv),
    \end{equation}
    where $g(x)=x\,{\rm coth}(x/2)-4$, $y$ is the so-called Compton-$y$ parameter (see below), and we have neglected all relativistic corrections (e.g.,~\cite{2006NCimB.121..487N}). The latter assumption is valid for our analysis because the void-tSZ cross-correlation is dominated by halos well below the mass scale for which relativistic corrections become significant (see Figure~\ref{fig:pe_mass}).
  
    The Compton-$y$ parameter associated to a particular structure at redshift $z$ is given by
    \begin{equation}\label{eq:sz_main}
      y(\theta)=\frac{\sigma_T}{m_e\,c^2}\int\frac{dr_\parallel}{1+z}P_e
      \left(\sqrt{r_\parallel^2+r_\perp^2}\right),
    \end{equation}
    where $\sigma_T$ is the Thomson scattering cross-section, $m_e c^2$ is the electron rest mass, $P_e(r)$ is the electron pressure profile of the structure, $r_\parallel$ and $r_\perp\equiv\chi(z)\theta$ are the longitudinal (parallel to the line of sight) and transverse comoving distances from the structure, $\chi(z)$ is the comoving distance to redshift $z$, and $\theta$ is the angular separation from the center of the projected structure.
    
    The tSZ signal around voids can therefore be predicted by estimating their expected excess electron pressure profile. This is directly connected to the problem of modeling the mechanisms by which baryons are heated in different environments, which has been approached from different angles in the literature. One approach is to assume that heating processes take place mostly in the dense environments of dark matter halos, and that the gas density and temperature can be related to halo mass (e.g., \cite{2001MNRAS.327.1353K}). Under this assumption, the void pressure profile can be directly computed in terms of two ingredients: the abundance of halos of different masses conditional to the environmental density, and a model for the relation between halo mass and gas density and pressure.
    
    Such a ``local'' heating mechanism would predict voids to be colder than the average, given the under-abundance of massive, hotter halos in underdense environments \cite{1996MNRAS.282..347M}. This description neglects other non-local sources of heating of the intergalactic medium (IGM), such as the effect of TeV blazars in the presence of plasma instabilities \cite{2012ApJ...752...23C}, which could even give rise to an inverted density-temperature relation \cite{2008MNRAS.386.1131B}.
    
    Here we will estimate the void tSZ signal by connecting the void density profile, which can be estimated directly from the data, with the electron pressure $P_e(r)$ through the so-called ``effective universe'' method. This approach is detailed in Appendix \ref{app:effu}, and has been previously used in analyses of environmental effects on halo abundances \cite{2003MNRAS.344..715G,2004ApJ...605....1G,2009MNRAS.394.2109M,2015MNRAS.447.2683A}). In short, one can associate the void underdensity $\delta(r)$ with a set of effective cosmological parameters $\Omega_X(r)$, which can then be used to estimate any quantity in the void as its background value in that effective cosmology.

    The problem of estimating the void pressure profile then reduces to computing the background free electron pressure for a given set of cosmological parameters. Assuming the main contribution to the total tSZ signal comes from the hot gas in dark matter halos (i.e., using the first ``local'' approach described above), the total electron pressure at a point ${\bf r}$ is given by the sum of the contributions from all halos:
    \begin{align}
      P_e({\bf r})=\int d^3{\bf x}\,dM\,n(M,{\bf x})P_e(|{\bf x}-{\bf r}|,M),
    \end{align}
    where $n(M,{\bf x})$ is the number density of halos of mass $M$ (i.e., the position-dependent halo mass function), with pressure profile $P_e(r,M)$. The background contribution to the electron pressure is therefore found by taking the ensemble average of the equation above:
    \begin{equation}\label{eq:pe_bg}
      \langle P_e \rangle=\int dM\,n(M)\frac{4\pi}{3}\int dr\,r^2\,P_e(r,M).
    \end{equation}
    Figure \ref{fig:pe_mass} shows the relative contribution to the background electron pressure in Eq. \ref{eq:pe_bg} as a function of halo mass at different distances from the void. As qualitatively expected, the contribution of massive halos is suppressed inside the void, and therefore this model predicts voids to be under-pressured.

    To summarize, the process to estimate the void's electron pressure profile is as follows:
    \begin{enumerate}
      \item Estimate the void's overdensity profile $\delta(r)$.
      \item At a given $r$, relate $\delta(r)$ to a set of effective cosmological parameters $\Omega_X(r)$ as described in Appendix \ref{app:effu}.
      \item The void's electron pressure at that $r$ is then computed using Eq. \ref{eq:pe_bg} as the background electron pressure for the corresponding effective cosmological parameters. 
      \item Integrate the void pressure profile along the line of sight (Eq. \ref{eq:sz_main}) to obtain the expected tSZ signal.
    \end{enumerate}
    Here we use the halo mass function of Ref.~\cite{2010ApJ...724..878T} and the electron pressure profile of Ref.~\cite{2012ApJ...758...75B}. The void density profile $\delta(r)$ is estimated directly from the data in terms of the galaxy overdensity (see Section \ref{ssec:results.densprof}). We compute the fiducial $y$ profile at a fixed redshift $z=0.5$, corresponding to the median redshift of the CMASS sample, and we verify that the resulting curve does not vary significantly with $z$ within the allowed redshift range. Note that, since all of our results are given in terms of the ratio $\theta/\theta_v$, where $\theta$ is the angular distance to the void center and $\theta_v$ is the projected void radius, redshift-dependent projection effects are negligible.

  \section{Data}\label{sec:data}
    \subsection{Void catalogs}\label{ssec:data.voids}
      We use the public void catalog described in Ref.~\cite{2017ApJ...835..161M}, constructed using the ZOBOV void finding algorithm \citep{2008MNRAS.386.2101N}, which connects underdensities identified through a Voronoi tessellation using a ``watershed'' method. The catalog is based on the 12th Data Release of the Baryon Oscillation Spectroscopic Survey (BOSS)\footnote{The BOSS data are available at \url{https://data.sdss.org/sas/dr12/boss/lss/}.} \cite{2016MNRAS.455.1553R}, part of the Sloan Digital Sky Survey. The full BOSS catalog, covering roughly 10,000 ${\rm deg}^2$, is sub-divided into two galaxy samples spanning complementary redshift ranges, LOWZ ($\sim4.6\times10^5$ objects, $0.2<z<0.43$) and CMASS ($\sim8.5\times10^5$ objects, $0.43<z<0.7$), and void catalogs are provided for both samples\footnote{The void catalogs used here are available at \url{http://lss.phy.vanderbilt.edu/voids/}.}. Although the authors identified more than 10,000 voids in the BOSS dataset, we focus our analysis only on the ``cut'' version of the catalogs, in which cuts on significance and minimum density were made to ensure a clean sample of truly underdense regions. In particular, we use the CMASS-based catalog, containing 774 voids. Each void is assigned an effective radius $r_v$ corresponding to the radius of the sphere encompassing its Voronoi volume. The median void size is $r_v\simeq34\,h^{-1}{\rm Mpc}$, subtending an angle of $\sim1.5^\circ$ at $z=0.5$.
      
      For this sample the authors also provide 1,000 mock realizations generated from a set of simplified N-body simulations. These mocks are based on a galaxy sample that reproduces the clustering properties of the CMASS sample, as well as its angular completeness and redshift distribution. The resulting void mock catalogs contain on average $\sim20\%$ more voids than the true data~\cite{2017ApJ...835..161M}, although they reproduce statistics of the true void data well in terms of angular, redshift, and size distribution. We therefore randomly downsample the mocks to correct for this issue. These mocks are used as random positions to estimate the null signal, and therefore to compute the stacked signal around the true voids. In addition, we use the mocks as an ensemble of random realizations to estimate the measurement uncertainties associated with the void stacking.
      
      In addition to the void catalogs, we also make use of the full CMASS galaxy sample, as well as the corresponding random catalogs made available by the BOSS Collaboration, to estimate the average void density profile.
      
    \subsection{tSZ and CMB maps}\label{ssec:data.cmb}
      In order to estimate the tSZ signal associated with voids we use the Compton-$y$ parameter maps made available by the \textit{Planck} Collaboration \cite{2016A&A...594A..22P}. The available maps were derived by applying internal linear combination (ILC) techniques to the \textit{Planck} intensity maps from 30 to 857 GHz. The \textit{Planck} Collaboration has released two $y$ maps derived using different reconstruction methods: the Modified Internal Linear Combination Algorithm (MILCA, \cite{2013A&A...558A.118H}) and the Needlet Internal Linear Combination (NILC, \cite{2011MNRAS.410.2481R}). In both cases, the separation of the tSZ signal from other sources of emission (CMB and foregrounds) is based mainly on its well-known frequency dependence, and both methods find linear combinations of the multi-frequency maps that minimize the variance of the resulting map while preserving a unit response to the tSZ frequency dependence and de-projecting the CMB. The methods also use spatial information by constructing independent weights for different scales and regions, although they differ in the details of how these weights are derived (see Ref.~\cite{2016A&A...594A..22P}).
      
      The NILC and MILCA maps have generally been found to give consistent results in various analyses (e.g.,~\cite{2016A&A...594A..22P,2017MNRAS.467.2315V,2017arXiv170603753H}), but the NILC map was found to have higher noise on large scales than MILCA~\cite{2016A&A...594A..22P}. This large-scale noise can be difficult to treat precisely in the absence of highly accurate random catalogs, especially when stacking on voids subtending relatively large angular scales.  Thus, we use MILCA as the fiducial $y$-map in this analysis, although we study the consistency of our results using the NILC map as well (see Section \ref{ssec:results.syst}).
      
      In order to mitigate the contamination from Galactic and extragalactic foregrounds, we use a combination of the \textit{Planck} 60\% Galactic mask and the union of the HFI and LFI point-source masks. In an effort to enhance the signal-to-noise ratio ($S/N$) of our measurement, we further mask all tSZ sources detected by \textit{Planck} \cite{2016A&A...594A..27P} above $5\sigma$ with redshifts $z<0.43$ (i.e., tSZ sources that are uncorrelated with the CMASS voids, which are by definition between $0.43<z<0.7$). We also make use of the HFI 545 GHz map \cite{2016A&A...594A...1P} to constrain the level of foreground and cosmic infrared background (CIB) contamination (see Section \ref{ssec:results.syst})\footnote{The \textit{Planck} data are available at \url{http://irsa.ipac.caltech.edu/Missions/planck.html}.}.

  \section{Results}\label{sec:results}
    \subsection{The void density profile}\label{ssec:results.densprof}
      Our prediction for the expected void $y$ profile (Section \ref{sec:theory}) requires an estimate of the average density profile. The void density profile $\delta(r)$ has been the subject of much study in recent years \cite{2002MNRAS.332..205A,2014MNRAS.440..601R,2016IAUS..308..542N}, and has been shown to take a fairly universal shape across size, redshift and, more importantly, tracer of the underlying density field \cite{2014PhRvL.112y1302H,2014MNRAS.442..462S}.
      
      We estimate the void underdensity from the density of tracer galaxies in the CMASS sample, using the corresponding random catalog to correct for edge effects and incompleteness. For each void $i$ we compute the number of data and random objects found in bins of $x=r/r^i_v$, where $r$ is comoving distance from the object to the center of the void and $r_v^i$ is the void's effective radius. The average density profile is then estimated as:
      \begin{equation}
        1+\delta_g(x)=\frac{N_R}{N_D}\frac{\sum_i D_i(x)}{\sum_i R_i(x)},
      \end{equation}
      where $D_i$ and $R_i$ are the distributions of data and random objects found around the $i^{\rm th}$ void, $N_D$ and $N_R$ are the total size of the data and random catalogs respectively, and the index $i$ runs over all the voids in the catalog.
      \begin{figure*}
        \centering
        \includegraphics[width=0.9\textwidth]{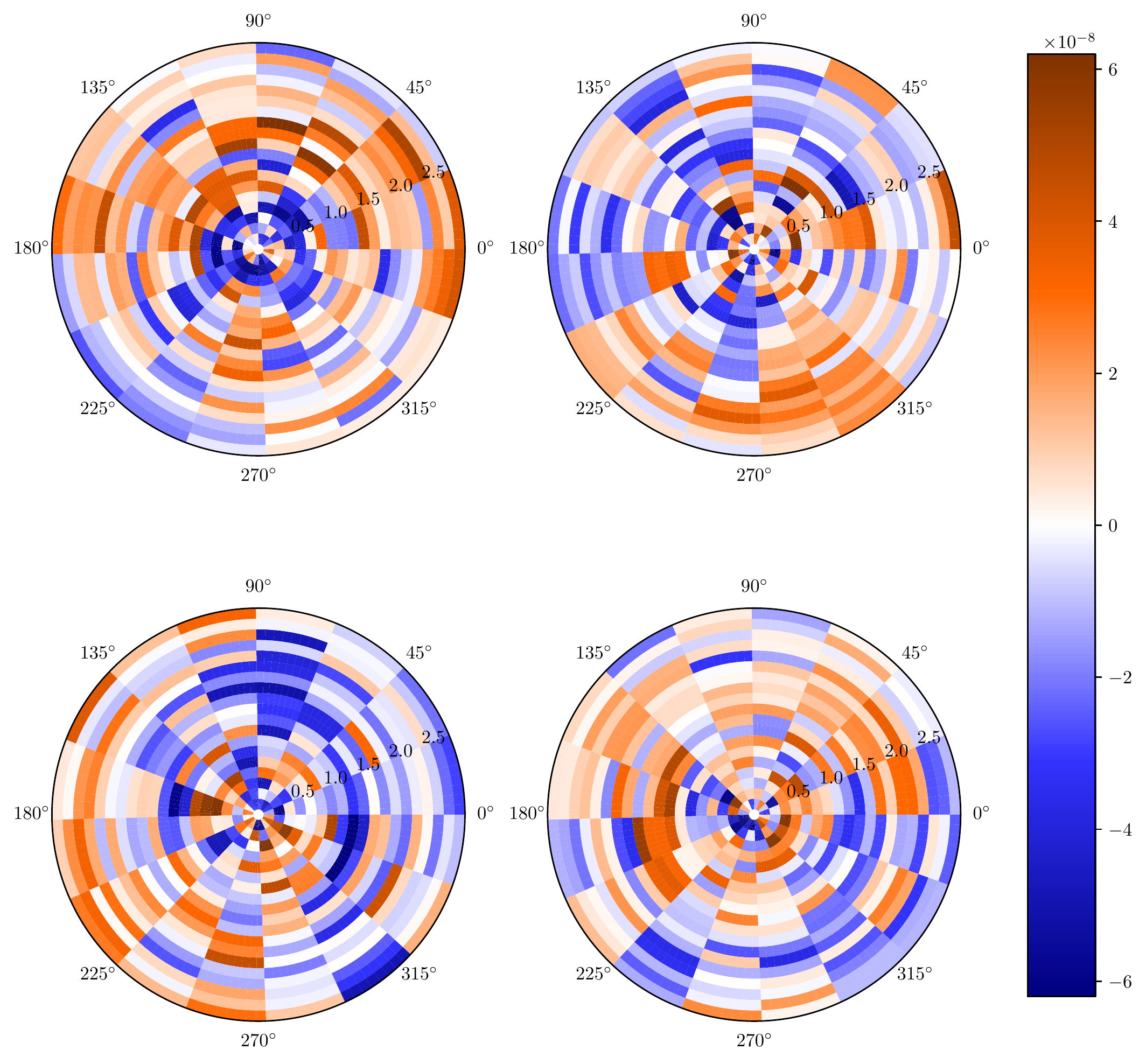}
        \caption{2-dimensional stacked tSZ signal in polar coordinates measured around the CMASS voids (top left) and three random mock void catalogs (top right and bottom). The data exhibit a noticeable average decrement below $\theta\lesssim0.7\theta_v$, where $\theta_v$ is the angle subtended by the void effective radius.}
        \label{fig:stacks_2d}
      \end{figure*}
      
      The density profile thus computed corresponds to the \textit{underdensity of tracer CMASS galaxies} around these voids at the median redshift $z\approx0.5$. The effective-universe approach, as described in Appendix \ref{app:effu}, is formulated in terms of the matter underdensity at redshift $z=0$. To compute this latter quantity in terms of $\delta_g(x|z=0.5)$ we must therefore account for the effects of galaxy bias and structure growth. To do so we simply rescale $\delta_g$ by a factor $[b_{\rm CMASS}D(z=0.5)]^{-1}$, where $b_{\rm CMASS}=2.0$ is the bias of the CMASS sample \cite{2013MNRAS.432..743N} and $D(z)$ is the linear growth factor normalized at $z=0$. Note that although in general non-linear contributions to both growth and galaxy biasing become important on small scales, recent studies find that this problem is alleviated around voids \cite{2017MNRAS.469..787P,2017JCAP...07..014H}, and this simple linear rescaling should be a good approximation given the uncertainties reported here.
      
    \subsection{The tSZ signal around voids}\label{ssec:results.yprof}
      In order to estimate the average tSZ signal around cosmic voids we proceed as follows:
      \begin{enumerate}
        \item For each void $i$ in the catalog, at redshift $z_i$ and with effective radius $r^i_v$, we loop over all pixels in the $y$ map lying within a radius $3\,\theta^i_v$ of the void's center, where $\theta^i_v=r^i_v/\chi(z_i)$ is the angle subtended by the void's effective radius. For each pixel $p$ we compute two quantities: $x_p\equiv \theta_{i,p}/\theta^i_v$ and $\psi_{p}$, where $\theta_{i,p}$ is the angular separation between the the pixel center and the void center, and $\psi_{i,p}$ is the angle that this separation vector forms with the great circle connecting the void center with the North Pole.
        \item For each void we then produce two 2-dimensional histograms, $s^i_y(x,\psi)$ and $s^i_N(x,\psi)$:
        \begin{align}\nonumber
          &s^i_y(x,\psi)=\sum_p \Theta(\psi_p,\psi,\Delta\psi)\Theta(x_p,x,\Delta x)\,y_p\\\nonumber
          &s^i_N(x,\psi)=\sum_p \Theta(\psi_p,\psi,\Delta\psi)\Theta(x_p,x,\Delta x),
        \end{align}
        where $y_p$ is the Compton-$y$ signal measured in pixel $p$, $\Theta(x_p,x,\Delta x)$ is a binning operator for a bin centered at $x$ with width $\Delta x$ (similarly for $\psi$) and $s^i_N(x,\psi)$ is the sum over positional (rather than tSZ) information only.
              
        We then estimate the average $y$ parameter in the $x$-$\psi$ plane for catalog $c$ as $\hat{y}_c(x,\psi)\equiv\sum_i s^i_y(x,\psi)/\sum_i s^i_N(x,\psi)$.
        \item We do this for the CMASS void catalog as well as the $N_{\rm m}=1000$ mock catalogs, and finally estimate the average tSZ signal corrected for sky coverage and completeness by subtracting the mock average:
        \begin{equation}\label{eq:stack_estimator}
          \bar{y}(x,\psi)=\hat{y}_{\rm CMASS}(x,\psi)-
          \frac{1}{N_{\rm m}}\sum_{c=1}^{N_{\rm m}}\hat{y}_c(x,\psi).
        \end{equation}
      \end{enumerate}
      In simpler terms, the estimator is therefore a simple stack around voids of the $y$ map in polar coordinates scaled by the effective void size. We have not implemented further refinements to the method, such as optimally filtering the $y$ map for each void as done in, e.g., Ref.~\cite{2017MNRAS.466.3364C}, which might marginally enhance the significance of this measurement, in order to facilitate the computation of the associated theoretical prediction. In our analysis we compute $\bar{y}(\theta/\theta_v,\psi)$ in 20 radial bins for $0\leq\theta/\theta_v\leq3$ and 16 angular bins for $0\leq\psi<2\pi$. We verify that the results presented here, in terms of both best-fit and detection significance, do not change when varying the radial sampling rate.
      \begin{figure}
        \centering
        \includegraphics[width=0.49\textwidth]{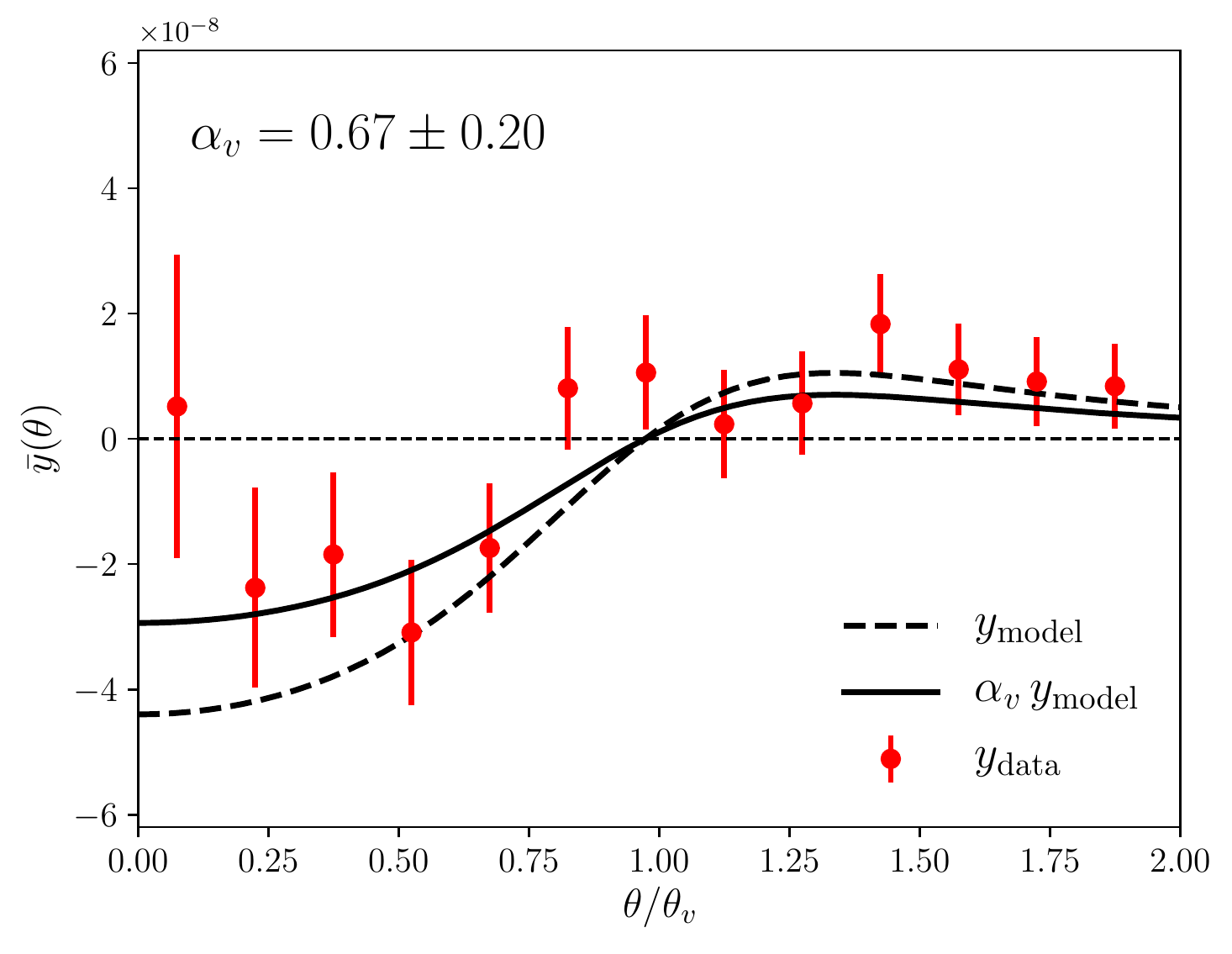}
        \caption{Radial tSZ profile around voids estimated from the CMASS void catalog (red circles with error bars). The dashed black line corresponds to the theoretical expectation based on the model described in Section \ref{sec:theory}, while the solid black line corresponds to the best-fit model found by scaling the fiducial prediction by an amplitude $\alpha_v$.}
        \label{fig:y_result}
      \end{figure}
      
      Figure \ref{fig:stacks_2d} shows the stacked tSZ signal around voids in polar coordinates for the CMASS catalog (upper left) and for three random mocks. Although the measurement is noisy, a consistent decrement in $y$ for $x<1$ with respect to the mean can be appreciated in the real data.
      
      Although the two-dimensional stacks are useful for visualization purposes, we do not expect a preferred orientation of the void signal, and therefore we proceed by considering only the radial tSZ profile (i.e., summing $s_y$ and $s_N$ over $\psi$) as our data vector. We further limit the size of this vector to the 13 $x$-bins with $x<2$, and write the profile measured in the $k$-th bin as $\bar{y}_k$.
      
      We estimate the covariance matrix of $\bar{y}_k$ from the scatter measured over the 1,000 mock catalogs:
      \begin{equation}
        C_{kk'}=\frac{1}{N_{\rm m}}\sum_{c=1}^{N_{\rm m}}
        \left(\bar{y}^c_k-\langle\bar{y}_k\rangle\right)\,
        \left(\bar{y}^c_k-\langle\bar{y}_k\rangle\right),
      \end{equation}
      where $\bar{y}^c_k$ is the measurement in the $c^{\rm th}$ mock, and $\langle\bar{y}_k\rangle$ is the average across mocks. This estimate of the covariance matrix was verified by an alternative computation of the diagonal errors via jackknife resampling. In order to quantify the significance of this measurement or the degree of agreement with a given model $y^{\rm mod}_k$, we compute the goodness-of-fit $\chi^2$:
      \begin{equation}
        \chi^2 \left(y^{\rm mod}\right)\equiv\sum_{k,k'}\left(\bar{y}_k-y^{\rm mod}_k\right)\,I_{kk'}\,\left(\bar{y}_{k'}-y^{\rm mod}_{k'}\right),
      \end{equation}
      where $I_{kk'}$ is the inverse covariance matrix. We estimate $I_{kk'}$ as the inverse of the sample covariance $C_{kk'}$ corrected for the overall scaling factor prescribed by \cite{2007A&A...464..399H}:
      \begin{equation}
        I_{kk'}=\frac{N_{\rm m}-n_d-2}{N_{\rm m}-1}\left(C^{-1}\right)_{kk'},
      \end{equation}
      where $n_d$ is the size of the data vector. We verified that the distribution of $\chi^2$ values for the 1,000 mock void catalogs for a null model ($y^{\rm mod}=0$, since the mocks and $y$ maps are uncorrelated) is well described by a ``chi-squared'' distribution with 13 degrees of freedom. Therefore, the $\chi^2$ value can be reliably interpreted as the likelihood of $y^{\rm mod}$ given the data $\bar{y}_k$.

      While a few models exist that can describe the heating (and therefore the tSZ signal) in voids, we remain agnostic about the particular form of any such model. Instead, we take a phenomenological approach and use the theoretical prediction described in Section \ref{sec:theory}, but allow for a rescaling amplitude $\alpha_v$. Since this is a simple linear parameter, the best-fit and standard deviation of $\alpha_v$ can be computed analytically as:
      \begin{align}
        \alpha_v&=\frac{\sum_{k,k'}y_k^{\rm mod}I_{kk'}\bar{y}_{k'}}
                       {\sum_{k,k'}y_k^{\rm mod}I_{kk'}y^{\rm mod}_{k'}},\\
        \sigma(\alpha_v)&=\left[\sum_{k,k'}y_k^{\rm mod}I_{kk'}y^{\rm mod}_{k'}\right]^{-1},
      \end{align}
      where $y^{\rm mod}$ is the theoretical model with a fiducial amplitude $\alpha_v^{\rm fid}=1$.
      \begin{figure}
        \centering
        \includegraphics[width=0.49\textwidth]{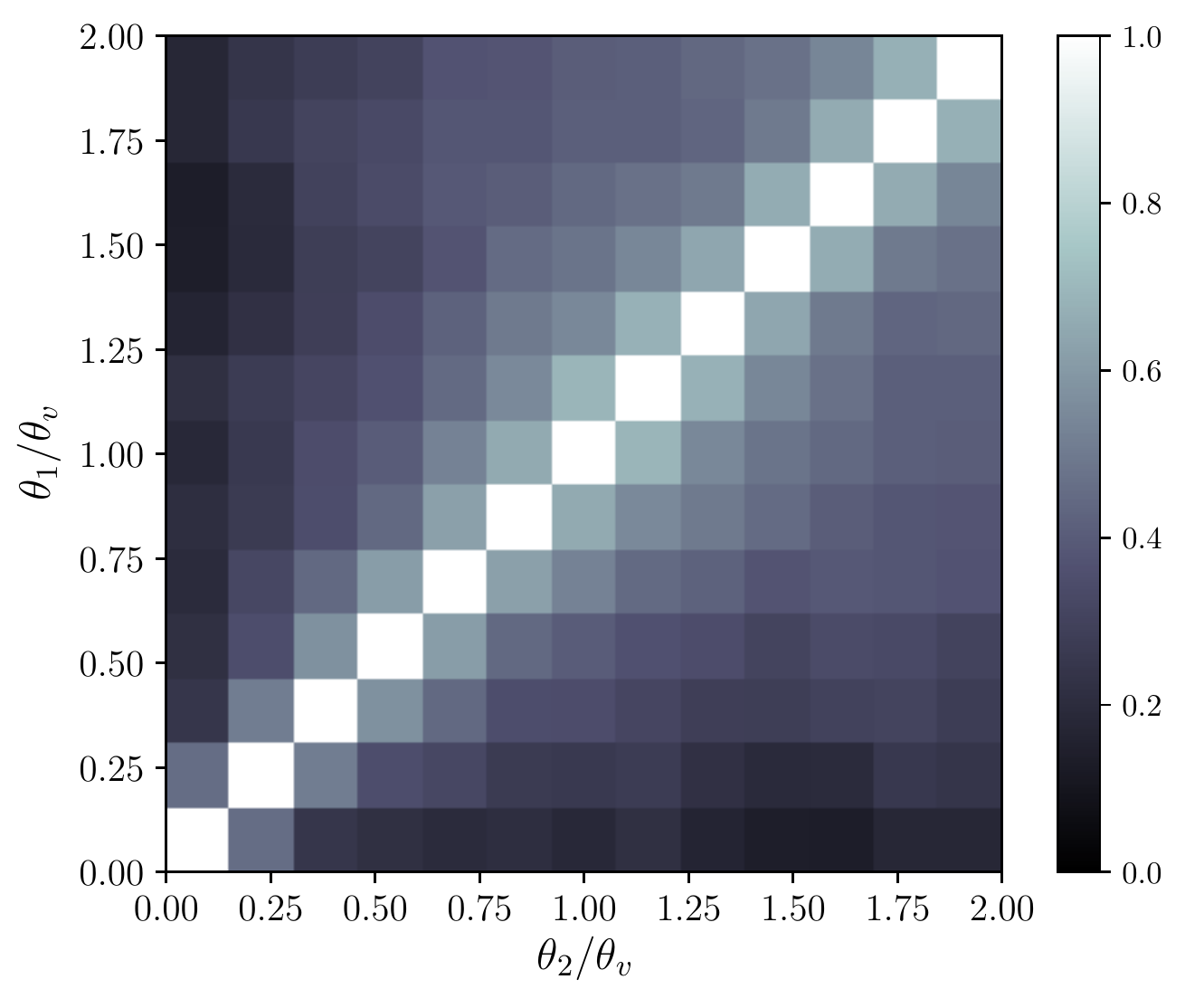}
        \caption{Covariance matrix of the measured radial tSZ profile around voids (see Figure \ref{fig:y_result}). The covariance matrix is estimated from the 1,000 CMASS mock void catalogues and contains significant off-diagonal elements that need to be accounted for in the analysis. These are caused by the beam smoothing of the $y$ maps and by the mixing of scales associated with the effective map rescaling with each void's size before stacking.}
        \label{fig:corrmat}
      \end{figure}
      
      We obtain a best-fit value and uncertainty on the rescaling amplitude
      \begin{equation}
        \alpha_v=0.668\pm0.199 \, (\mathit{Planck}\,\mathrm{MILCA}\,y\mathrm{-map}).
      \end{equation}
      This corresponds to a $3.4\sigma$ measurement of the tSZ signal associated with cosmic voids. Figure \ref{fig:y_result} shows the measured signal (red circles with error bars), the fiducial theory prediction (dashed black line), and the best-fit scaled model (solid black line). The $\chi^2$ for this best-fit model is $\chi^2(\alpha_v)=15.3$, corresponding to a probability-to-exceed (PTE) of 0.22 for 12 degrees of freedom. In contrast, for the null model we obtain $\chi^2({\rm null})=26.6$, with a PTE of 0.012. The significance of this measurement in terms of a $\chi^2$-difference is therefore $\sqrt{\Delta\chi^2}=3.36$, in agreement with our previous estimate.
      
      Figure \ref{fig:corrmat} shows the estimated correlation matrix $R_{kk'}\equiv C_{kk'}/\sqrt{C_{kk}C_{k'k'}}$. Note that because of the beam smoothing of the $y$ maps, as well as the mixing of scales caused by the effective rescaling of the map with void size, there are significant off-diagonal contributions to the covariance, which need to be accounted for.
      
    \subsection{Null tests and systematics}\label{ssec:results.syst}
      In order to test the robustness of this measurement, we consider the possible
      impact of certain systematic uncertainties and perform a number of consistency tests.
      
      A first simple test for the presence of systematic errors is cross-correlating the CMASS voids with the ``half-difference'' Compton-$y$ map, corresponding to the difference of the $y$ maps constructed using the first and second halves of stable pointing periods, and distributed together with the full MILCA map. The half-difference map should therefore contain only noise and no real $y$ signal (or other astrophysical signals). We carry out the same analysis described in Section \ref{ssec:results.yprof} on this half-difference map, including the computation of the associated covariance matrix, and find that the signal measured from the data is compatible with zero, with $\chi^2=12.8$ (${\rm PTE}=0.464$).
      
      We also verify the consistency of our measurement by repeating it on the NILC Compton-$y$ map. We find that the measured void $y$ profile agrees with the measurement from the MILCA map up to an overall additive offset. As mentioned in Section \ref{ssec:data.cmb} and pointed out by Ref.~\cite{2017MNRAS.467.2315V}, the NILC map suffers from a higher large-scale noise power than the MILCA map (see Fig. 5 in Ref.~\cite{2016A&A...594A..22P}). This large-scale contribution, as well as the overall amplitude of the void signal, are much smaller ($\sim O(10^{-8})$) than the typical per-pixel noise ($\sim O(10^{-6})$). Therefore any imperfection in the removal of the mean contribution to the correlation function estimator (i.e., the second term on the right hand side of Eq. \ref{eq:stack_estimator}), such as small deviations in the mock void catalogs from the true BOSS footprint, may give rise to an overall offset in the estimator. This is particularly relevant for the void stacks, given the larger angular scales involved, compared to the usual stacking analyses around groups or clusters.
      \begin{figure}
        \centering
        \includegraphics[width=0.49\textwidth]{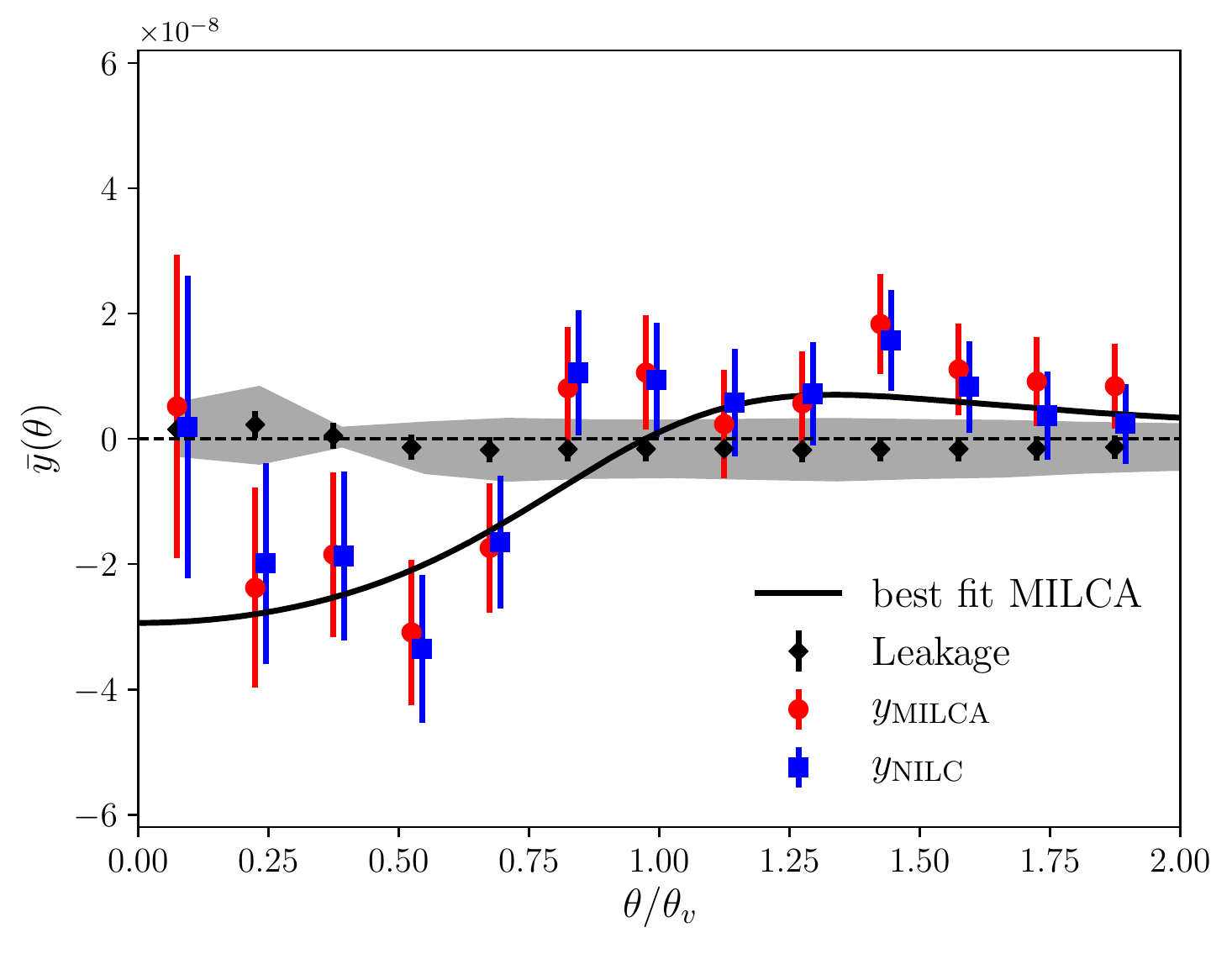}
        \caption{Radial tSZ profile around voids estimated from the CMASS void catalog using the MILCA $y$ map (red circles with error bars) and the NILC $y$ map (blue squares with error bars). The solid black line corresponds to the best-fit rescaling of the theoretical expectation described in Section to the MILCA map. The black diamonds with error bars show the best-fit dust leakage computed by scaling the stacked void signal on the \textit{Planck} 545 GHz map by a constant factor $\alpha_{\rm CIB}$ estimated as described in Section \ref{ssec:results.syst}. The gray band around these data shows the dust leakage allowed by the $1\sigma$ uncertainty on $\alpha_{\rm CIB}$.}
        \label{fig:y_syst}
      \end{figure}

      To correct for this issue, we introduce an extra free parameter in our model, corresponding to an overall additive amplitude $\alpha_{\rm off}$, and fit for it jointly with the amplitude of the void profile $\alpha_v$. The measurement of the void $y$ profile in the NILC map corrected for this offset is shown as blue squares in Fig. \ref{fig:y_syst}, which also shows the original MILCA measurement in red. This procedure yields a measurement of $\alpha_v$ from the NILC map that is in agreement with our previous estimate, $\alpha_v= 0.64\pm 0.21$, with a similar significance. The measured offset $\alpha_{\rm off}= (1.4\pm 0.5)\times10^{-8}$ is significant at the $2.8\sigma$ level, and the overall fit is good, with a ${\rm PTE}=0.21$. It is also worth pointing out that, after repeating this analysis on the MILCA map, we find that the measured offset is compatible with $0$, and that the recovered value of $\alpha_v$ and its uncertainty do not change significantly with respect to the fiducial analysis.
      
      Finally, we quantify the level of contamination of our measurement by other potential correlated components. In particular, we focus on the contribution from imperfectly cleaned extragalactic dust emission (CIB), which is a known contaminant of the \textit{Planck} $y$ maps \cite{2016A&A...594A..22P,2016A&A...594A..23P}\footnote{Since the CMB component was explicitly deprojected in the construction of both the MILCA and NILC maps, our measurement is immune to any contamination from the void integrated Sachs-Wolfe signal.}. As a first step, we follow the procedure described in \cite{2014JCAP...02..030H, 2017MNRAS.467.2315V}, making use of the \textit{Planck} 545 GHz map as a tracer of dust emission. We outline the method here, and we refer the reader to Refs.~\cite{2014JCAP...02..030H,2017MNRAS.467.2315V} for further details.
      
      We start by assuming that the $y$ map is contaminated by CIB and Galactic dust emission, such that the observed map is
      \begin{equation}
        y_{\rm obs}=y_{\rm true}+\alpha_{\rm CIB}\,T_{\rm CIB}+\alpha_{\rm Gal}T_{\rm Gal},
      \end{equation}
      and that the 545 GHz map is dominated by precisely these components
      \begin{equation}
        T_{\rm 545}=T_{\rm CIB}+T_{\rm Gal}.
      \end{equation}
      We can then determine the leakage amplitudes $\alpha_{\rm CIB}$ and $\alpha_{\rm Gal}$ by analyzing the auto-correlation of the 545 GHz map and its cross-correlation with the observed $y$ map. This also requires the use of existing models for the CIB power spectrum and its true cross-correlation with the tSZ signal, for which we use the measurements of \cite{2014A&A...571A..30P} and \cite{2016A&A...594A..23P}, respectively. After masking 80\% of the sky we obtain $\alpha_{\rm CIB}=(2.3\pm6.6)\times10^{-7}\,({\rm MJy/sr})^{-1}$ and $\alpha_{\rm Gal}=(-0.8\pm1.9)\times10^{-7}\,({\rm MJy/sr})^{-1}$.
      
      Since the Galactic component should not correlate with the void distribution (unless regions of large Galactic dust absorption could affect the void finding procedure), the most dangerous source of leakage is the CIB component. The contribution of this source of contamination to the measured $y$ void profile, $\bar{y}_{545}$, can therefore be quantified by repeating the void stacking measurement on the 545 GHz map and scaling the resulting signal, $\bar{T}_{545}$ with $\alpha_{\rm CIB}$: $\bar{y}_{545}=\alpha_{\rm CIB}\,\bar{T}_{545}$. The resulting estimated leakage is shown as black diamonds in Figure \ref{fig:y_syst}, with the shaded region corresponding to the level of leakage allowed by the $1\sigma$ uncertainties on $\alpha_{\rm CIB}$. The conclusion is that the leakage is generally much smaller than the measured tSZ signal.
      
      \begin{figure}
        \centering
        \includegraphics[width=0.49\textwidth]{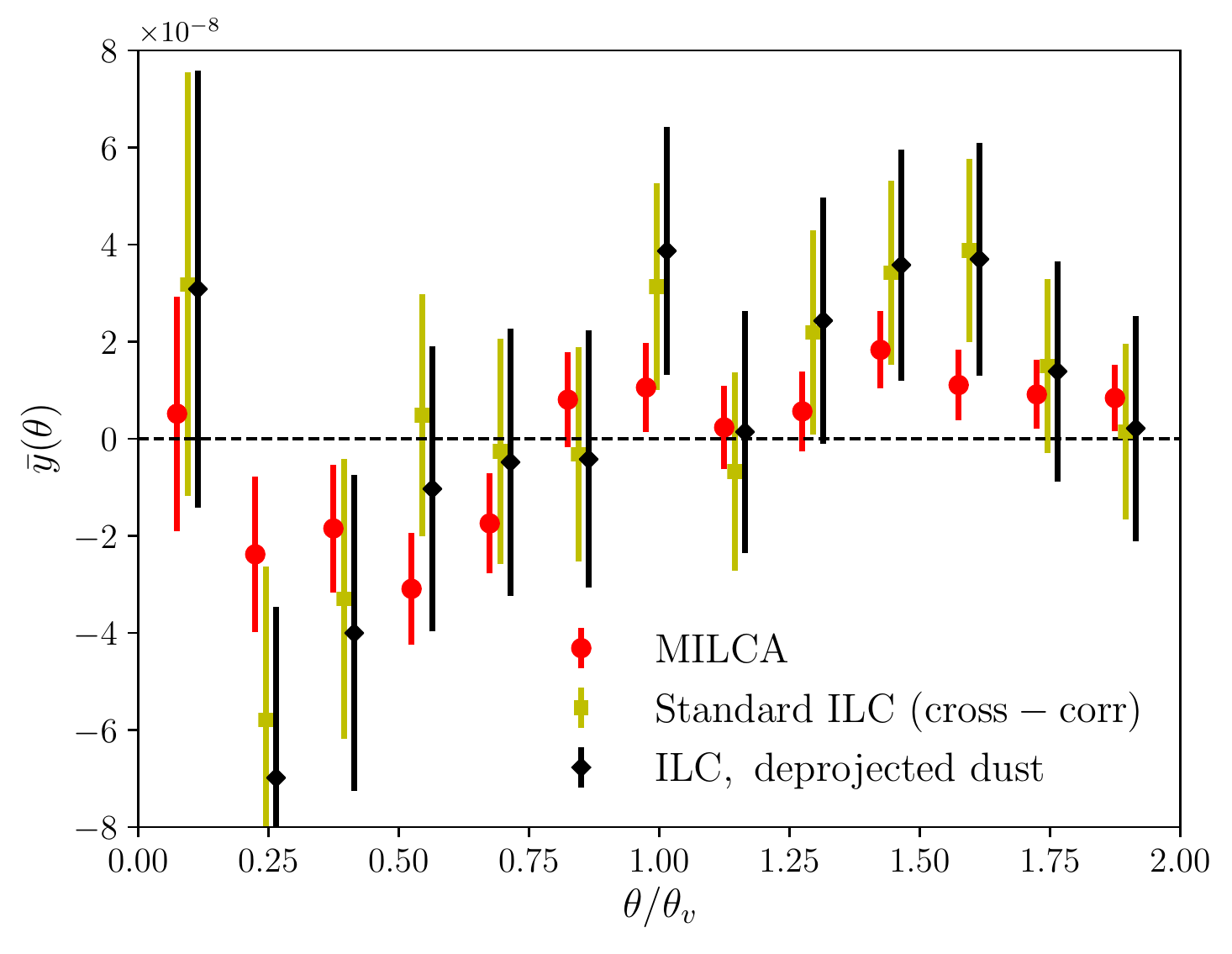}
        \caption{Radial tSZ profile around voids estimated from the CMASS void catalog using the MILCA $y$ map (red circles with error bars) via our fiducial analysis, and estimated by applying an ILC at the cross-correlation level to measurements of the void cross-correlation with the six \emph{Planck} HFI frequency maps. The yellow squares show the result for a standard (tSZ-preserving, variance-minimizing) ILC applied to the cross-correlation measurements, while the black diamonds show the result when the ILC additionally deprojects a fiducial CIB component.  The consistency of the yellow and black points indicate that the ILC is already removing CIB contamination effectively. Although the direct HFI - void cross-correlation measurements are noisier than our fiducial results, they demonstrate robustness to possible CIB contamination.}
        \label{fig:y_ilc}
      \end{figure}
      However, there is an important assumption in this method for assessing the CIB leakage, which unfortunately is not strictly valid for the MILCA or NILC $y$ maps.  In particular, the method assumes that the power spectrum of the CIB leakage into the $y$ map can be treated as an overall amplitude multiplying the true CIB power spectrum.\footnote{For the $y$ map constructed in Ref.~\cite{2014JCAP...02..030H}, the ILC weights were scale-independent, so this assumption was valid.} In the MILCA and NILC $y$ maps, the varying filters used as a function of multipole in the reconstructions lead to scale-dependent ILC weights.  While the tSZ power spectrum is preserved by MILCA/NILC, the power spectrum of contaminating components can have a scale-dependence that differs strongly from their true physical shape (see, e.g., Fig.~14 in Ref.~\cite{2016A&A...594A..23P}, where the CIB leakage in the $y$ map auto-spectrum has a much steeper shape than the true CIB power spectrum). The upshot is that one cannot self-consistently assess the CIB leakage by cross-correlating the MILCA/NILC $y$ maps with the 545 GHz map and fitting a CIB power spectrum model.
      
      Thus, although the leakage estimated is already small, we consider an additional method to demonstrate robustness.  In particular, we directly measure the cross-correlation of the \emph{Planck} HFI maps (100 - 857 GHz) with the void catalog, and implement multi-frequency foreground cleaning at the level of the cross-correlation.  The cross-correlation measurement pipeline is identical to that described in Section \ref{ssec:results.yprof}. We combine the six measurements (one for each HFI frequency) using an ILC applied to the cross-correlation results themselves (rather than at the map level).  The ILC weights preserve the tSZ signal and minimize the variance of the resulting linear combination. We additionally consider a ``constrained ILC'' that also deprojects a fiducial CIB component corresponding to the best-fit modified-blackbody spectrum of \cite{2013JCAP...07..025D}. The results are shown in Figure~\ref{fig:y_ilc} and compared to the results of our fiducial analysis. Although the error bars increase, the results are consistent.  Moreover, the very small changes seen when imposing the CIB deprojection in the cross-correlation ILC demonstrate that the method is already removing CIB contamination effectively.
      
      Finally, the approximate level of CIB contamination is also confirmed by a rougher estimate of $\alpha_{\rm CIB}$, given by the ratio of the cross-correlation between the $y$ and the 545 GHz maps to the auto-correlation of the latter ($\alpha_{\rm CIB}=C_\ell^{y\times545}/C_\ell^{545\times545}\lesssim3\times10^{-7}\,({\rm MJy/sr})^{-1}$). The agreement amongst several methods for assessing the CIB contamination provides confidence that our result is not dominated by this systematic.  Nevertheless, higher-significance measurements may require a more detailed analysis to ensure that small amounts of CIB leakage do not lead to a bias.

  \section{Discussion}\label{sec:discussion}
    Cosmic voids have proven to be a useful tool for cosmological analyses. Linear perturbation theory holds in a wider range of scales around them, and they allow us to explore structure formation in low-density environments, dominated by vacuum energy and populated by lower-mass halos. Voids also allow us, through the cross-correlation with maps of the tSZ effect, to explore the presence of hot baryons in underdense regions, and to put constraints on different models for the heating of the IGM.
    
    Here we have presented the first stacking analysis of tSZ maps released by the
    \textit{Planck} Collaboration on voids detected in the CMASS sample of the BOSS survey. To quantify the significance of the detection of this cross-correlation we have fit the measured stacked $y$ profile to the theoretical prediction described in Section \ref{sec:theory} scaled by an overall free amplitude $\alpha_v$. We find $\alpha_v=0.67\pm0.20$, a $3.4\sigma$ detection of void underdensities in the Compton-$y$ maps. We have verified that this measurement is robust against null tests, contamination of the $y$ maps by dust/CIB, and the choice of component separation method. For the sake of reproducibility we make our full analysis pipeline available at \href{https://github.com/damonge/VoidSZ}{https://github.com/damonge/VoidSZ}.
    
    While larger and more sensitive datasets will be needed to increase the sensitivity of this detection and confirm it, some qualitative conclusions can already be extracted. First of all, the gas in underdense regions probed by voids is also under-pressured, as predicted by the simple model used here, based on the modulation of halo abundances in environments of different densities. However, although our measurement is $3.4\sigma$ away from the null case, it is also $1.6\sigma$ away from the negative amplitude predicted by this model, implying that voids could be warmer than one might naively expect. We may however speculate on the reasons for this marginal tension.
    
    Our theoretical model is arguably imprecise: on the one hand, even though the effective-universe method is an exact result at the background level (at least for spherical underdensities), it fails at predicting the growth of perturbations and can therefore lead to a mis-estimation of the abundance of halos in voids \cite{2012PDU.....1...24A}. This is more generally related to the problem of modeling the conditional mass function, which has proven to be difficult to do precisely \cite{2008MNRAS.383..546C}. On the other hand, our model also uses an estimate of the halo pressure profile extrapolated to the lower-mass halos that dominate the tSZ signal inside the void --- the pressure profile model was originally calibrated only for halos with mass $\gtrsim 5 \times 10^{13} \, M_{\odot}/h$~\cite{2012ApJ...758...75B}. An upturn in the $y$ -- mass relation toward low masses could therefore explain the departure from $\alpha_v=1$, although evidence from tSZ -- galaxy group cross-correlation measurements does not favor this explanation~\cite{2013A&A...557A..52P,2015ApJ...808..151G,2015MNRAS.451.3868L,2017arXiv170603753H}. However, it is possible that the pressure content of low-mass groups in voids systematically differs from that of similar-mass groups in average-density environments. Future simulation analyses may shed light on such environmental effects.
    
    A more interesting possibility would be the presence of non-local heating mechanisms, such as the effects of TeV blazars on the IGM advocated by Ref.~\cite{2012ApJ...752...23C}. These models predict an inverted temperature-density relation, generating warmer voids than we would otherwise expect. Our measurement then suggests that a more comprehensive study of tSZ stacks on environments of different densities could be an effective way to put constraints on these models.

    Analyses like the one presented here will, in the future, benefit from larger-volume galaxy surveys like those conducted with the Dark Energy Spectroscopic Instrument \cite{2013arXiv1308.0847L},  the Large Synoptic Survey Telescope \cite{2009arXiv0912.0201L} or the Wide Field Infrared Survey Telescope \cite{2015arXiv150303757S}. The higher number density of tracers in these datasets will also improve the robustness of the associated void catalogs and allow a more optimal measurement against the deepest underdensities. In addition, these measurements will only improve in significance with large-area, high-resolution maps of the tSZ effect through ground-based multi-frequency experiments like the Advanced Atacama Cosmology Telescope \cite{2014JCAP...08..010C,2016SPIE.9910E..14D} and the South Pole Telescope Third Generation instrument \cite{2014SPIE.9153E..1PB}, as well as the upcoming Simons Observatory\footnote{\url{http://www.simonsobservatory.org}} and, looking to the more distant future, CMB-Stage IV \cite{2016arXiv161002743A}, given adequate cleaning of other sources of contamination. Similar analyses could also be carried out with these datasets to detect the kinematic SZ signal of voids, probing the properties of dark energy through the velocity profiles of these objects.

\section*{Acknowledgements}
  The authors would like to thank Ariel Amaral, Nick Battaglia, Bhuvnesh Jain, Adam Lidz, Mark Richardson, Ben Wandelt, and Hans Winther for useful comments and discussions. DA and RH also thank the Center for Computational Astrophysics, part of the Flatiron Institute, for their hospitality. The Simons Foundation supports the Flatiron Institute. DA acknowledges support from the Science and Technology Facilities Council and the Leverhulme and Beecroft Trusts. This work was partially supported by a Junior Fellow award from the Simons Foundation to JCH. The Dunlap Institute is funded through an endowment established by the David Dunlap family and the University of Toronto. RH would like to acknowledge that the land on which the University of Toronto is built is the traditional territory of the Haudenosaunee, and most recently, the territory of the Mississaugas of the New Credit First Nation. The territory was the subject of the Dish With One Spoon Wampum Belt Covenant, an agreement between the Iroquois Confederacy and the Ojibwe and allied nations to peaceably share and care for the resources around the Great Lakes. This territory is also covered by the Upper Canada Treaties. RH is grateful to have the opportunity to work in the community, on this territory.
  
\bibliography{paper}

\appendix
\section{The effective-universe approach to void-related quantities}\label{app:effu}
  It is a well-known result, valid in both Newtonian and relativistic gravitational theory (e.g., \cite{1947MNRAS.107..410B}), that a spherically symmetric overdensity residing in an otherwise homogeneous Universe will evolve, at any distance $r$ from its center, as a parallel Friedmann-Lema\^itre-Robertson-Walker universe with effective cosmological parameters \cite{Weinberg:1972,Peebles:1993}. These can be related to the density profile $\delta(r)$ and local infall velocity of the overdensity (the latter defining the local expansion rate) as:
  \begin{align}
    &\Omega_M(r)=\Omega_M^{\rm BG}\frac{1+\Delta(r)}{\eta^2(r)},\hspace{6pt}
    \Omega_\Lambda(r)=\frac{\Omega_\Lambda^{\rm BG}}{\eta^2(r)},\\
    &H_0(r)=H_0^{\rm BG}\eta(r),
    \hspace{6pt}\Delta(r)\equiv\frac{3}{r^3}\int_0^rds\,s^2\,\delta(s),
  \end{align}
  where $\Delta(r) $ is the average overdensity enclosed within a sphere of radius $r$, $\eta(r)$ is proportional to the local infall velocity normalized by the background expansion, and all quantities labelled ${\rm BG}$ are the cosmological parameters of the background universe. The ratio between expansion rates can be fixed by imposing a homogeneous age of the Universe:
  \begin{equation}
    t_{\rm BB}=\frac{1}{H_0}\int_0^1
    \frac{dx}{x\sqrt{\Omega_Mx^{-3}+\Omega_\Lambda+\Omega_Kx^{-2}}},
  \end{equation}
  effectively making the perturbation a purely growing mode that vanishes at early times.
  
  The computation of the background tSZ signal (Eq. \ref{eq:pe_bg}) requires an estimate of the halo mass function, which depends on the evolution of both the background and linear perturbations. For deep underdensities, the cosmological constant's contribution to the total energy density dominates over that of matter, and therefore perturbations grow \textit{more slowly at late times than in the background cosmology}. This effect can be taken into account by scaling the value of $\sigma_8$ outside the void by the ratio of the linear growth factors in the effective and background cosmologies with the same normalization at early times.

\end{document}